\documentstyle[buckow1,12pt]{article}
\begin {document}
\def\titleline{
The role of the Polyakov loop in the Dirac 
\newtitleline
operator of QCD at finite temperature}
\def\authors{Vicente Azcoiti}
\def\addresses{Departamento de Fisica Te\'orica, Universidad de Zaragoza, \\
Plaza San Francisco, 50009 Zaragoza, Spain}
\def\abstracttext{
We show how all the contributions to the determinant of the 
Dirac-Kogut-Susskind operator of QCD at finite temperature 
containing a net number of Polyakov loops become irrelevant in 
the infinite volume limit. We discuss also on two of the most 
interesting physical implications of this result: i) the restoration 
of the Polyakov symmetry in the full theory with dynamical fermions 
and ii) the total suppression of baryonic thermal fluctuations in 
QCD at finite temperature.}
\makefront
\section{Introduction}
The fermion determinant in QCD and in general in any gauge theory is 
a gauge invariant non local operator, the main contributions of which 
being complex combinations of closed Wilson loops including the Wilson 
line or Polyakov loop \cite{POLY} 
which is closed trough the boundary of the finite 
time direction.

In the pure gauge model the gauge action is invariant under Polyakov 
transformations and, because of that, the mean value of the Polyakov 
loop has been extensively used as order parameter in the investigations 
of the finite temperature phase transition. The situation however change 
in the full theory with dynamical fermions where the Polyakov symmetry 
is explicitly broken by the contribution to the integration measure 
of the determinant of the Dirac operator. The general wisdom in this case 
is that the Polyakov loop is no more an order parameter for the finite 
temperature phase transition.

Notwithstanding that I will show how in lattice QCD with staggered fermions 
all the contributions to the fermion determinant containing a net 
number of Polyakov loops become irrelevant in the infinite spatial volume 
limit. The Polyakov symmetry is recovered in this limit and we can therefore 
kill all these contributions from the beginning and work with a theory with 
dynamical fermions which preserves the Polyakov symmetry. 

But this is, maybe, neither the only nor the most interesting physical 
implication of this result. In fact the Hilbert space of physical states 
can be decomposed as a direct sum of Hilbert spaces, each one of them 
corresponding to a fixed value of the quark number operator. The conservation 
of baryonic charge in QCD implies that all these spaces are invariant 
under the hamiltonian and we can therefore write the partition function 
as a sum of canonical partition functions, each one of them corresponding 
to a fixed value of the quark number operator. This decomposition of the 
partition function coincides with the integrated partition function 
obtained from the Polyakov loop expansion of the fermion determinant, 
the quark content being the net number of Polyakov loops. Since, as we 
will show, all the contributions with a net number of Polyakov loops 
are irrelevant, we conclude that thermal fluctuations of physical states 
with non vanishing baryonic charge are fully suppressed in QCD at finite 
temperature.

A method to implement in standard simulations the Polyakov symmetry and 
to kill baryonic thermal fluctuations in full QCD will also be developed 
here. The rest of the paper is organized as follows: section 2 describes 
some general features of the partition function of QCD at finite temperature. 
In section 3 we show the main result of this work with the help of an extra 
abelian degree of freedom. Section 4 is devoted to discuss some interesting 
physical consequences which follow from the results of section 3. In section 
5 we discuss on the generalization of these results for Wilson fermions.

\section{The Partition Function}
The partition function of QCD at finite temperature T

\begin{equation}
Z=Tr(e^{-H\over{T}})
\label{1}
\end{equation}
is the trace over the Hilbert space of physical states of minus the inverse 
temperature times the hamiltonian. Taking into account the conservation of 
baryonic charge we can write the partition function as a sum of canonical 
partition functions at fixed baryon number as follows:

\begin{equation}
Z=\sum_{k}Tr_{k}(e^{-H\over{T}}),
\label{2}
\end{equation}
where $Tr_k$ in (2) indicates the trace over the subspace of fixed quark 
number k and $k=0, +3, -3, +6, -6,...$

The decomposition of the partition function given in (2) is the standard 
representation used in the investigations of QCD at finite baryon density. 
In this last case the partition function (2) is slightly modified by the 
introduction of a chemical potential $\mu$ in the following way \cite{KOG}, 
\cite{HAS} 

\begin{equation}
Z=\sum_{k}e^{{\mu\over{T}}k}Tr_{k}(e^{-H\over{T}}).
\label{3}
\end{equation}

The previous decomposition (2) of the partition function as a sum of 
canonical partition functions at fixed baryon number corresponds in 
lattice regularized QCD to the Polyakov loop expansion of the integrated 
partition function which, for staggered fermions, can be written as follows

\begin{equation}
Z=\bar{a}_{3V_x}+\bar{a}_{3(V_x-1)}+.....
+\bar{a}_{0}+\bar{a}_{-3}+.......+\bar{a}_{-3V_x},
\label{4}
\end{equation}
where $\bar{a}_i$ $(\bar{a}_{-i})$ is the integral over the gauge group of all 
the contributions to the fermion determinant containing i forward (backward) 
Polyakov loops respectively, the integral being weighted with the pure gauge 
Boltzmann factor. The maximum number of Polyakov loops which can appear in a 
given coefficient of (4) for SU(3) and Kogut-Susskind fermions is three times 
the spatial lattice volume. $\bar{a}_0$ is the contribution with no net number 
of Polyakov loops and of course we have also all the symmetric contributions 
corresponding to backward Polyakov loops.

Since each coefficient in (4) represents the partition function at a fixed 
baryon number, this expression is also consistent with the fact that $V_x$ 
is the maximum number of baryons allowed by the Fermi-Dirac statistics in a 
finite and discrete space of $V_x$ points and staggered fermions.

All the averaged coefficients $\bar{a}_i$ in (4) are positive definite since 
they correspond to the decomposition of the partition function as a sum of 
canonical partition functions over the subspaces of fixed baryon number. Due 
to the Z(3) Polyakov symmetry of the pure gauge action, the only non 
vanishing coefficients are those containing a multiple of three times Wilson 
lines, which on the other hand reflects the fact that the only physical 
states in QCD are mesons, baryons and combinations of them.

\section{The Extra-Abelian Degree of Freedom}
Let us consider the determinant of the Dirac-Kogut-Susskind operator for the 
slightly modified gauge configuration which consists in multiplying all link 
variables at a fixed time-slice and pointing forward in the time direction 
by the global phase factor $e^{i\eta}$ and all the hermitian conjugates by  
$e^{-i\eta}$. Taking into account that the standard way to introduce a chemical 
potential in the lattice \cite{HAS} 
is to multiply all links pointing forward (backward) 
in the time direction by a factor $e^{\mu}$ $(e^{-\mu})$ respectively, this is 
just what corresponds to consider an imaginary chemical potential 
$\mu = i\eta/L_t$, $L_t$ being the lattice temporal extent.

The fermion determinant for a given gauge configuration and with the previous 
extra-degree of freedom can be written as follows

\begin{equation}
Det \Delta(i\eta)= a_{3V_x}e^{i3V_x\eta} + 
.......+a_1 e^{i\eta}+ a_0+a_{-1}e^{-i\eta}+
.......+a_{-3V_{x}}e^{-i3V_x\eta}.
\label{5}
\end{equation}
The coefficients $a_i$ $(a_{-i})$ in (5), averaged over the gauge group 
with the corresponding pure gauge Boltzmann factor, are the coefficients 
$\bar{a}_i$ $(\bar{a}_{-i})$ which appear in (4).

\begin{equation}
\bar{a}_{i}= \int [DU] a_{i}(U) e^{-\beta S_{G}(U)}.
\label{6}
\end{equation}

The first interesting property of expression (5) which follows from the 
hermiticity and chiral properties of the fermion matrix $\Delta$ is that 

\begin{equation}
Det \Delta(i\eta)\geq{0}
\label{7}
\end{equation}
for every $\eta$.

By inverse Fourier transformation we can write

\begin{equation}
a_j = {1\over{2\pi}}\int e^{-ij\eta} det \Delta(i\eta) d\eta , 
\label{8}
\end{equation}
$a_j=0$ if $j>{3V_x}$, $j<{-3V_x}$. These relations and the 
inequality (7) together 
tell us that $a_0\geq{0}$ and the absolute values $|a_j|\leq{a_0}$ 
for every j. 
Furthermore the same relations also hold for the integrated coefficients which 
appear in (4), i.e. 

\begin{equation}
\bar{a}_k \leq \bar{a}_0 , 
\label{9}
\end{equation}
where the absolute value in (9) disappears because of the fact that the 
integrated coefficients are real and positive.

Taking into account the symmetry properties of the coefficients 
$(\bar{a}_k = \bar{a}_{-k})$ we can write the partition function (4) as follows

\begin{equation}
Z = \bar{a}_0 (1+ {{2\bar{a}_3}\over{\bar{a}_0}}+
...........+ {{2\bar{a}_{3V_x}}\over{\bar{a}_0}}), 
\label{10}
\end{equation}
which gives for the free energy density $f$ the following expression

\begin{equation}
f = {T\over{V_x}} log \bar{a}_0 + 
{T\over{V_x}} log ( 1+ {{2\bar{a}_3}\over{\bar{a}_0}}+
...........+ {{2\bar{a}_{3V_x}}\over{\bar{a}_0}}).
\label{11}
\end{equation}
This expression and the inequalities (9) imply that in the thermodynamical 
limit $V_x\rightarrow\infty$ the free energy density can be computed as:

\begin{equation}
f = {T\over{V_x}} log \bar{a}_0, 
\label{12}
\end{equation}
i.e., the only relevant contribution in the determinant of the Dirac operator 
to the thermodynamics of QCD at finite temperature is that corresponding to a 
zero net number of Polyakov loops. We can therefore kill all the irrelevant 
contributions from the beginning and restore the Polyakov symmetry in full QCD. 
\section{Some Relevant Physical Implications}
The main result of the analysis here developed is contained in equation (12).
This section will be devoted to discuss some physical 
consequences which follow from it.

i) Since the only relevant contribution to the partition function of QCD at 
finite temperature is that with zero net number of Polyakov loops, the 
Polyakov symmetry is restored in the infinite volume limit of full QCD with 
dynamical fermions. The Polyakov loop is therefore a good order parameter 
for full QCD. A practical way to implement this result 
in numerical simulations is 
to include an extra-abelian degree of freedom, as done in section 3 of this 
paper.

ii) Another interesting conclusion which follows from the fact that the 
coefficient $\bar{a}_0$ in (12) does not depend on the boundary conditions 
for the fermion field is that periodic and antiperiodic boundary conditions 
give rise to the same physics.

iii) The physical meaning of equation (12) for the free energy is that the 
partition function of QCD at finite temperature is dominated by the 
canonical partition function computed over the Hilbert subspace of physical 
states of vanishing baryon number. In other words, baryonic thermal 
fluctuations are fully suppressed in QCD at finite temperature.

iv) It has been pointed out recently that in numerical simulations of 
quenched QCD at finite temperature and in the broken deconfined phase, 
the chiral condensate seems to depend crucially on the $Z_3$ phase in 
which the gauge dynamics settles \cite{UNO} and the chiral 
symmetry restoration 
transition appears to occur at different temperatures depending of the phase 
of the Polyakov loop \cite{MISHA}. After the analysis here 
developed it is clear that 
the correct way to solve this puzzle and to implement the quenched 
approximation in QCD is to take for the chiral condensate operator its 
Polyakov loop invariant part. This result also suggest that a investigation 
of the finite size effects in finite temperature full QCD induced by the 
irrelevant contributions to the partition function could be of great interest.
\section{Wilson Fermions}
We have shown in the previous sections of this paper how the thermodynamics 
of QCD, when regularized in a space-time lattice and using staggered fermions, 
is controlled by the contribution to the fermion determinant with no net number 
of Polyakov loops, i.e., by the thermal fluctuations of physical states with 
vanishing baryon number.

The two main ingredients to get this result are the conservation of 
baryonic charge in QCD and the inequalities (9) of section 3 which tell us 
that the partition function at fixed baryon number reaches its maximum value 
in the Hilbert subspace corresponding to zero baryon number. The first one 
of the two ingredients is independent of the lattice regularization for the 
Dirac operator. The second one however is based on the positivity of the 
determinant of the Dirac operator for any gauge configuration and any value 
of the extra-abelian degree of freedom $e^{i\eta}$ (equation (7) of section 3).

Since we have made use of the hermiticity and chiral properties of the 
Dirac-Kogut-Susskind operator in order to get equation (7), it is natural 
to ask whether our result applies to any fermion regularization or rather 
it is related to the presence-absence of the chiral anomaly.

Let us say from the beginning that even if we have not yet a definite 
answer to this question, there are strong indications suggesting that the 
chiral anomaly does not play any relevant role here. These indications come 
from the analysis of the properties of the fermion determinant for 
Wilson fermions. As well known, the Dirac-Wilson operator $\Delta$ can 
be written as 

\begin{equation}
\Delta = I - \kappa M, 
\label{13}
\end{equation}
where $\kappa$ is the hopping parameter and the matrix M verifies the 
following chiral relation

\begin{equation}
\gamma_{5} M \gamma_{5} = M^{+}. 
\label{14}
\end{equation}

Equation (14) implies that if $\lambda$ is eigenvalue of M, $\lambda^{*}$ is 
also eigenvalue of M, i.e., the fermion determinant is always real. However 
it could be negative and in fact this unpleasant situation has been found 
for some gauge configurations in numerical simulations of the Schwinger model 
done in the unphysical strong coupling region \cite{NOS}. However the unitary 
character of the gauge group implies that all the eigenvalues are upper 
bounded by the relation

\begin{equation}
|\lambda | \leq{8} 
\label{15}
\end{equation}
which implies that for $\kappa \leq{1/8}$, $det \Delta \geq{0}$. It is easy 
to verify that under the previous condition $\kappa \leq{1/8}$, the 
positivity of $det \Delta$ also holds in the presence of the extra-abelian 
degree of freedom introduced in section 3.

In other words, all the results of this paper can be extended in a 
straightforward way to Wilson fermions if we impose the restriction 
$\kappa \leq{1/8}$, i.e., the hopping parameter region associated to a 
positive bare fermion mass.

\section{Aknowledgements}
This work has been partially supported by CICYT (Proyecto AEN97-1680). 
The author thanks Giuseppe Di Carlo and Angelo Galante for discussions.


\end{document}